# Modeling interstellar amorphous solid water grains by tight-binding based methods: comparison between GFN-XTB and CCSD(T) results for water clusters

Aurèle Germain[1] and Piero Ugliengo[1,2][0000-0001-8886-9832]

[1] Dipartimento di Chimica, Università degli Studi di Torino, via P. Giuria 7, 10125, Torino, Italy.
[2] Nanostructured Interfaces and Surfaces (NIS) Centre, Università degli Studi di Torino, via P. Giuria 7, 10125, Torino, Italy.

**Abstract.** One believed path to Interstellar Complexes Organic Molecules (iCOMs) formation inside the Interstellar Medium (ISM) is through chemical recombination at the surface of amorphous solid water (ASW) mantle covering the silicate-based core of the interstellar grains. The study of these iCOMs formation and their binding energy to the ASW, using computational chemistry, depends strongly on the ASW models used, as different models may exhibit sites with different adsorbing features. ASW extended models are rare in the literature because large sizes require very large computational resources when quantum mechanical methods based on DFT are used. To circumvent this problem, we propose to use the newly developed GFN-xTB Semi-empirical Quantum Mechanical (SQM) methods from the Grimme's group. These methods are, at least, two orders of magnitude faster than conventional DFT, only require modest central memory, and in this paper we aim to benchmark their accuracy against rigorous and resource hungry quantum mechanical methods. We focused on 38 water structures studied by MP2 and CCSD(T) approaches comparing energetic and structures with three levels of GFN-xTB parametrization (GFN0, GFN1, GFN2) methods. The extremely good results obtained at the very cheap GFN-xTB level for both water cluster structures and energetic paved the way towards the modeling of very large AWS models of astrochemical interest.

## 1 Introduction

The interstellar medium (ISM) contains a vast diversity of complex organic molecules (COMs) [1], but the formation of these interstellar COMs (iCOMs) is still a mystery for the most part. The chemical reactions producing iCOMs could be done in the gas phase of the ISM, but some observed chemical species were proven to be unstable when produced in the gas phase [2]. A way for them to be stabilized would be by the way of a third body in which the reaction energy can be injected [3]. This third body can be the silicate dust core grains in diffuse clouds or, in dense molecular clouds (MC), the multiple layers of amorphous water ice (AWS) covering dust core [4].

These water layers are not formed by direct adsorption of water molecules but by H and O recombination, first at the grain core surface, and then on top of the pre-formed water layers [5]. The molecules present in the gas phase of dense MC highly influence the composition of the layers, and so the ice water is "dirtied" by other molecular species like carbon monoxide, carbon dioxide, methanol, and ammonia to name a few [4]. It is assumed that these chemical species, after being adsorbed on the surface of these icy grains, will form iCOMs and desorb from the grains to return in the gas phase. A lot of unanswered questions arise, mainly involving the binding energies of these molecular species to the ice mantle, the way they diffuse from site to site, and by which mechanism they eventually desorb. Astronomical observations give us limited answers to these questions, and the extreme conditions of the ISM (low temperature and low densities of the species studied), coupled with the time frame in which these reactions are believed to occur, are nearly impossible to emulate in a laboratory. Computational chemistry gives us a way to circumvent these limitations and study what is impossible to observe or difficult to reproduce in terrestrial labs. Nonetheless, these studies are highly influenced by the way we model the AWS ice mantles.

Usually, in the literature, the AWS models are represented by water clusters envisaging very few molecules. Clearly, these models cannot represent neither the structural complexity nor the hydrogen bond cooperativity exhibited by the ice in large AWS grains. The paucity of the models is also due to our ignorance about how to model them within a physically sound approach, *i.e.* obeying the rules of their formation in the ISM. Therefore, the development of realistic AWS models is a key factor in astrochemistry, as they will be essential to compute accurate binding energies of the various iCOMs adsorbed on the grains. Unfortunately, the high computational power demanded to produce large enough AWS model grain can grow steeply with the cluster size, *de facto* preventing the application of current DFT methods to model large icy particles. The strategy adopted in our group is based on models defined within the periodic boundary conditions, either through crystalline ice unit cell or amorphous ones on which the results of the cheap HF-3c [11] method were compared with that of the more accurate B3LYP-D3 method [10]. The inclusion of the periodic boundary conditions mitigates the smallness of the unit cell size, better mimicking an extended AWS. On those models we have computed the binding energies of about 20 iCOMs, showing a relatively large dependence of these values on the adsorption site. The problem is that increasing the size of the unit cell to increase the variability of the surface sites is hampered by the cost of the calculation, even at the cheapest HF-3c level. Therefore, a different strategy is needed to really deal with large AWS models based on even cheaper methods than HF-3c to cope with larger sizes. However, cheaper methods may be inaccurate to the point in which the predictions become unreliable and, therefore, they should be carefully benchmarked against accurate quantum mechanical results.

In this paper, we benchmark the accuracy of a new family of semiempirical quantum mechanical (SQM) methods, the GFN-xTB methods (GFN-xTB stands for Geometry, Frequency, Noncovalent, eXtended Tight Binding) developed by Grimme and coworkers [6][7][8]. GFN-xTB methods allow pushing the limit of the system size up to hundreds of molecules (or thousands atoms), and have been shown to be at least two order of magnitude faster than conventional DFT methods [6], while keeping a good level of accuracy for a variety of molecular properties, including intermolecular interactions, usually not well accounted for by SQM. The three different levels of GFN-xTB (GFN0, GFN1, GFN2) envisage different level of parametrization and treatment of dispersion interactions, providing different accuracy. In the following, we check for the accuracy of all of them using water clusters as a reference system as they exhibit similar features of larger ASW clusters.

## 2  Water cluster study

### 2.1  Energetic features

We focused on a set of 38, already optimized, water clusters by the work of Temelso et al. 2011 [9]. These clusters are ranging from two to ten water molecules each. Geometries were optimized up to RI-MP2/aVDZ level and then, since the difference in geometry is minute between the two [9], a single-point calculation was performed using large basis-set and CCSD(T)/CBS/CBSnocp (Coupled Cluster Single Double Triple within the complete basis set extrapolation). We re-optimized these water clusters at GFN-xTB with the three levels of parametrization, by using a single laptop in less than half a day of computing time. On the optimized geometries, we computed the normalized binding energy ($BE_N$) of each water cluster as:

$$BE_N = \frac{BE}{N_{H_2O}} = \frac{E_{H_2O} \times N_{H_2O} - E_{WC}}{N_{H_2O}}$$

Where $BE$ is the non-normalized binding energy, $E_{WC}$ is the total energy of a water cluster of $N_{H_2O}$ water molecules and $E_{H_2O}$ is the total energy of a fully optimized isolated water molecule. The BE value includes, therefore, the cost of geometry re-organization of water molecules caused by the building up of the cluster.

On Fig. 1 we contrasted the BE energies computed by GFN-xTB against the ones calculated by Temelso et al. CCSD(T)/CBS/CBSnocp level as a function of water cluster nuclearity. The black colored line of Fig. 1 is a best fit line through the 38 points while, the grey line is the "ideal" trendline that would represent the perfect agreement between GFN-xTB and CCSD(T) BE values.

Looking at Fig. 1, it is easy to notice that the first level of theory shown with GFN0 is not very accurate, while both GFN1 and GFN2 showed trendlines very close to the ideal one. Comparison between GFN1 and GFN2 also showed a better internal linear correlation coefficient R for GFN2 compared to GFN1.

The good accuracy of the results was expected from the general GFN-xTB benchmark work already conducted by Grimme [6][7][8], but it is impressive to see this level of agreement with CCSD(T) data, considering the negligible computer time requested for GFN-xTB compared to CCSD(T).

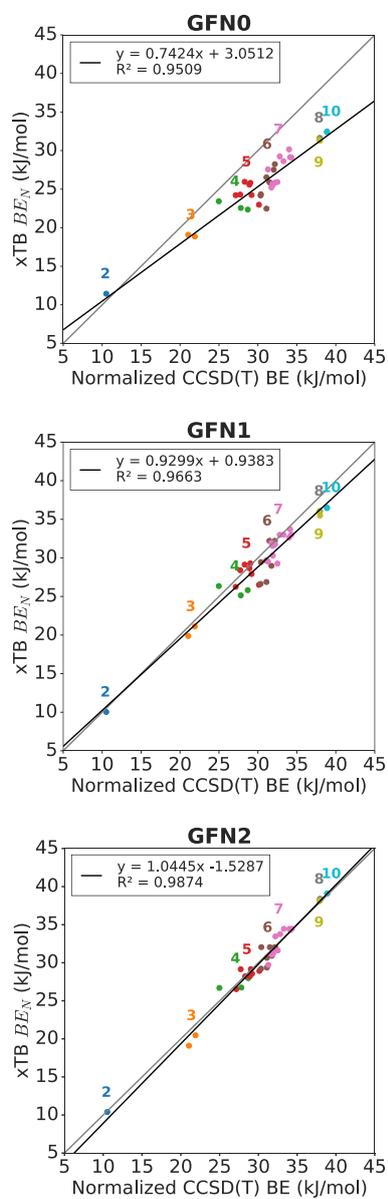

**Fig. 1.** Comparison between normalized BE values of the 38 water clusters computed by GFN-xTB and CCSD(T). GFN0, GFN1, GFN2, refer to the level of GFN-xTB parametrization (see text for details). Labels indicate the nuclearity of each water cluster.

Fig. 2 shows the absolute percentage of the difference (APD) between the BE computed at CCSD(T) and those at GFN-xTB level, defined as:

$$APD = \left|\frac{BE(CCSD(T)) - BE(GFN-xTB)}{BE(CCSD(T))}\right| \times 100$$

Results of Fig. 2, showed similar trend in the accuracy of the different GFN-xTB parametrization level, with errors of the GFN2 being almost half of those for GFN1 and GFN0, both providing exceedingly large deviations. An interesting point is that the APD decreases dramatically as the system size increases, a result extremely relevant for the forthcoming simulation of large AWS grains. We refrain from a too definitive conclusion about this matter, since we only have two water clusters envisaging 8, 9, and 10 water molecules, but it is very promising for our future research.

In conclusion, for the energetic part, we have showed GFN-xTB in its GFN2 incarnation to show a minimum/maximum/average APD values of about 1%/9%/3%, respectively. In the next paragraph we will discuss the structural features predicted by GFN-xTB against the accurate ones.

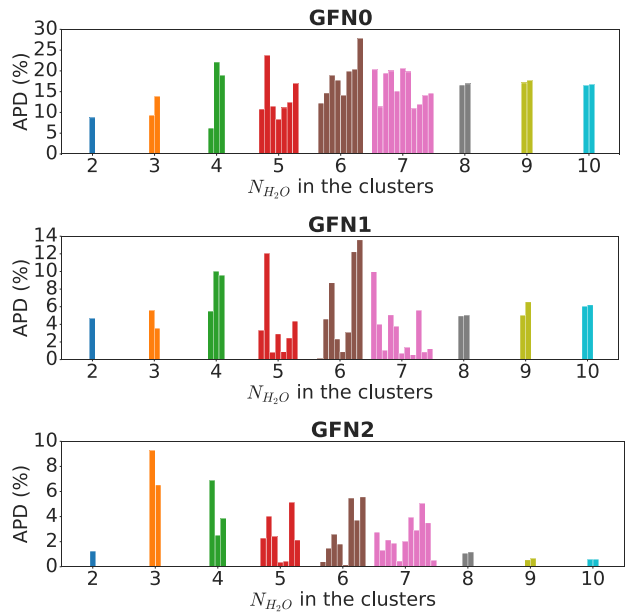

**Fig. 2.** APD between BE GFN-xTB and CCSD(T). Note that the boundaries of the y-axis change between the three plots.

### 2.2 Structural features

As we are dealing with a large number of water cluster, a one by one comparison of their geometrical features will be too cumbersome. Therefore, we relied on a global indicator of geometrical similarity by adopting the root mean squared deviation (RMSD) as a figure of merit (the smaller the RMSD the better) between the GFN-xTB-optimized geometries and the reference ones. The RMSD are shown in Fig. 3, arranged by the number of water molecules of each cluster. The data shows, for a given cluster nuclearity, different RMSD values as a result of the different water organization of each cluster. For instance, clusters with N=7 exhibit many different water configurations and, therefore, each of them shows a different RMSD value. We see here a slightly different behavior of the GFNx levels compared to the energetic results, as GFN0 provides RMSD values less spread than GFN1. GFN2 is, again, the most reliable method, while all of them give excellent results for clusters of nuclearity greater than 7. This is related to the much-reduced configuration space of water molecules for clusters at higher nuclearity compared to the smaller ones. A visual comparison limited to the water cluster with N=7 is showed in Fig. 4. GFN2 provides almost a perfect match with the high-level structure, while GFN0 is giving a RMSD almost half than that of GFN1. This is confirmed by the average RMSD values of 0.30/0.38/0.36 Å for GFN2/GFN1/GFN0, respectively.

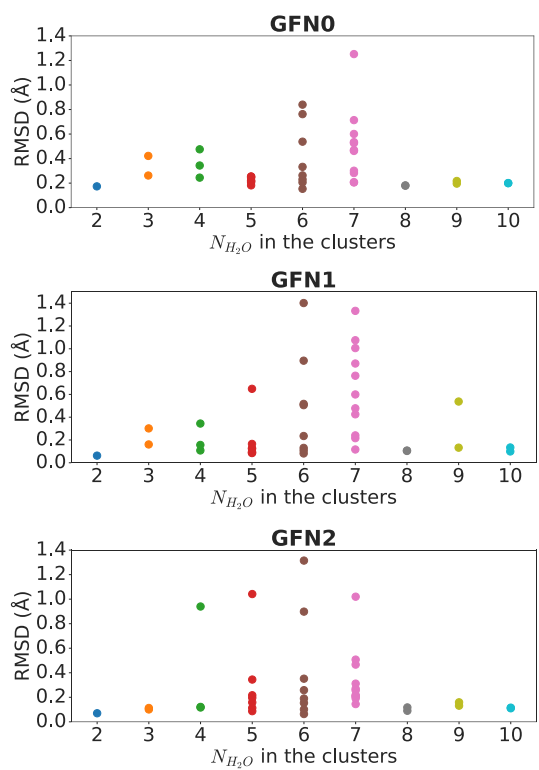

**Fig. 3.** RMSD for the water clusters. Different points for each cluster nuclearity represent different water organization within the cluster itself.

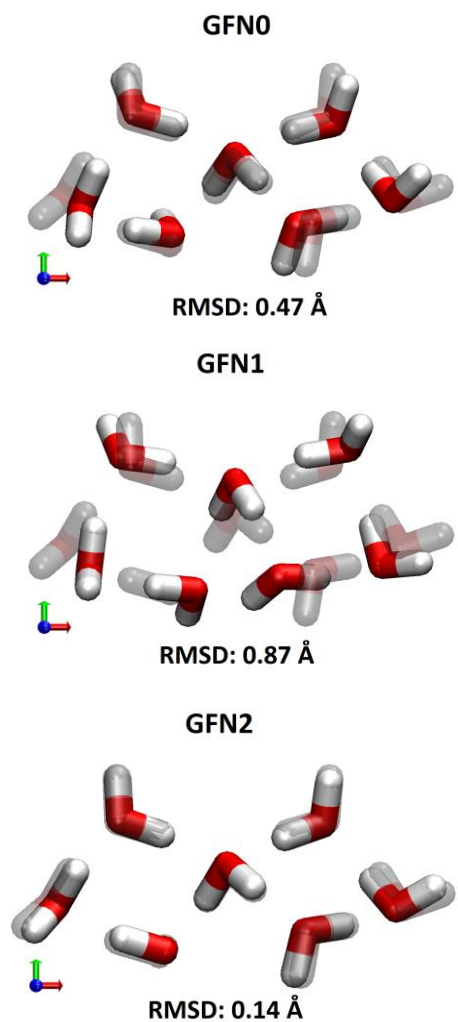

**Fig. 4.** Superposition of the GFN structures (full body) with the ones obtained at high level of theory (fade body) for one configuration of the water cluster exhibiting seven water molecules. RMSD values in unit of Å.

## 3   Conclusions

In this paper we have studied the accuracy of the new tight binding methods GFN-xTB recently proposed by Grimme and coworkers [6][7][8] to simulate the energetic and structures of water clusters. The reason for focusing on water clusters is that they are prototypes of AWS interstellar grains, as they share the same kind of intermolecular interactions (*i.e.* hydrogen-bond and dispersion interactions) and their size is lim-ited enough to be treated at high level of theory to be served as a reference benchmark for the GFN-xTB methods.

Here we used the energetic and structural results for 38 water clusters already studied in the literature [9] which were compared to three different flavors of GFN-xTB methods, namely GFN0, GFN1, and GFN2, corresponding to different level of parametrization. Our results showed that GFN2 is the most accurate and reliable GFN-xTB incarnation, for both structures and energetics. For structure, the average RMSD for the whole set of water cluster was 0.30/0.38/0.36 Å for GFN2/GFN1/GFN0, respectively, while the whole average APD in the BE were 3/5/16% for GFN2/GFN1/GFN0. We observe that both RMSD and APD in the BE are much smaller for large cluster nuclearity, paving the way to treat very large water cluster reminiscent of real AWS dust with good accuracy. This will certainly be possible due to the extreme computational efficiency of the GFN-xTB approach compared to the standard DFT one or even more so for the most expensive post-Hartree-Fock methods. We are already working in our laboratory to extend the size of the cluster in order to compare the BE of iCOMs species at the ice grain surfaces with that of our recent work on periodic water ice models [10].

## Acknowledgements

This project has received funding from the European Union's Horizon 2020 research and innovation programme under the Marie Skłodowska-Curie grant agreement No 811312 for the project "Astro-Chemical Origins" (ACO).

Fig. 4 was made with VMD. VMD is developed with NIH support by the Theoretical and Computational Biophysics group at the Beckman Institute, University of Illinois at Urbana-Champaign.